\documentstyle[preprint,aps,epsfig]{revtex}
\draft
\tighten

\begin{document}

\title{Dissipative collisions in $^{16}$O + $^{27}$Al at E$_{lab}$=116
MeV }


\author{  C.  Bhattacharya, K. Mullick, S. Bhattacharya, K.Krishan,
T. Bhattacharjee, P. Das,\\
 S. R. Banerjee, D. N. Basu, A. Ray,  S. K. Basu and
M. B. Chatterjee$^*$}

\address{  Variable Energy Cyclotron Centre, 1/AF Bidhan Nagar, Kolkata -
700 064, India\\ $^* $Saha Institute  of  Nuclear  Physics,  1/AF  Bidhan
Nagar, Kolkata - 700 064, India}

\maketitle

\begin{abstract}

The  inclusive  energy  distributions  of  fragments  (  3$\leq$Z$\leq$7)
emitted in the reaction $^{16}$O + $^{27}$Al at $E_{lab} = $116 MeV  have
been  measured  in  the  angular  range  $\theta_{lab}  $=  15$^\circ$  -
115$^\circ$. A non-linear optimisation procedure using multiple  Gaussian
distribution  functions  has  been proposed to extract the fusion-fission
and  deep  inelastic  components  of  the  fragment  emission  from   the
experimental  data.  The  angular  distributions  of  the fragments, thus
obtained,  from the deep inelastic component are found to fall off faster
than  those  from  the  fusion-fission component, indicating shorter life
times  of  the  emitting  di-nuclear  systems.  The  life  times  of  the
intermediate  di-nuclear  configurations  have  been  estimated  using  a
diffractive Regge-pole model. The life times thus extracted  ($\sim  1  -
5\times  10^{-22}$  Sec. ) are found to decrease with the increase in the
fragment charge.  Optimum  Q-values  are  also  found  to  increase  with
increasing charge transfer i.e. with the decrease in fragment charge.

\end{abstract}


The  phenomenon  of  fragment  emission  in  light heavy-ion collision at
energies $\leq$ 10 MeV/u has evolved a lot of interest  in  recent  years
(\cite{Sanders99}  and references therein). The origin of these fragments
extends        from         quasi-elastic         (QE)/projectile-breakup
\cite{Carlin89,Paladino90},  deep  inelastic  (DI)  transfer and orbiting
\cite{shapi2,shiva,shapi82,cormier,Mikumo80,Sato83,Szanto97},          to
fusion-fission                                                       (FF)
\cite{Sanders99,Szanto97,Sanders91,Matsuse97,dha2,bhat95,bhat96}
processes and in some cases the structure of the nuclei has been found to
play  an  important  role.  The  distinction  between  different reaction
mechanisms, in general, and the DI and FF processes,  in  particular,  is
very     difficult     for    light    system    (A$_{cn}    \leq    40$)
\cite{Sanders99,Szanto97} as in these cases there is  strong  overlap  in
the elemental distributions of the fragment emitted in the two processes.
The DI components are characterised by large energy damping and the fully
damped  yields,  in  general,  correspond  to  FF  components. The energy
damping observed in DI processes is due to the manifestation  of  nuclear
viscosity.  Thus, by a systematic study of DI fragments it is possible to
extract  information  on  the  nuclear  viscosity  parameters  which  are
important  for  understanding  nuclear  fission dynamics. This is usually
accomplished by studying the systematics of optimum  Q  values  vs.  mass
transfer and angle of rotation of the dinuclear complex. Thus, it is very
much  essential  to decipher the data to extract the contribution of each
component ( e.g. DI, FF) present in the  fragment  emission  spectra,  in
order to understand the underlying reaction dynamics.

Several  studies made earlier for $^{16}$O + $^{27}$Al system at incident
energies in the range of $\sim$60 -- 100 MeV have
indicated  that  fragments emitted in the reaction are mainly originating
from cluster  transfer  \cite{Harris77},  projectile  sequential  breakup
\cite{Carlin89} and multi-nucleon transfer \cite{cormier,Mikumo80,Sato83}
processes.  The role of direct two-body and three-body projectile breakup
in fragment emission from $^{16}$O + $^{27}$Al reaction in the energy
range of $\sim$70 -- 125 MeV have also been investigated recently
\cite{Paladino90}.  However,  none  of the earlier workers did attempt to
estimate the contribution of fusion-fission as a  competing  process  for
fragment   emission   in  $^{16}$O  +  $^{27}$Al  reaction.  It  is  well
established,    both    theoretically     \cite{Szanto97,moretto}     and
experimentally  (e.g. \cite{Sanders99}), that for systems lying below the
Businaro-Gallone point, asymmetric fission of the compound  nucleus  (CN)
contributes  significantly  in  the  fragment  emission  scenario. In the
present work we have studied  the  fragment  emission  spectra  from  the
reaction $^{16}$O + $^{27}$Al at $E_{lab} = 116$ MeV and report here, for
the  first  time,  a  simple  prescription  to  extract the FF and the DI
components of the fragments yield following the decay of light  composite
systems (A$_{cn} \leq 43$).

The  experiment  was  performed  using $116$ MeV $^{16}$O$^{5+}$ ion beam
from the  Variable  Energy  Cyclotron  at  Kolkata,  which  was  recently
upgraded  with  electron  cyclotron resonance (ECR) heavy ion source. The
target  used  was  420  $\mu$g/cm$^2$  self-supporting   $^{27}$Al.   The
fragments  were  detected  using  three  solid state(Si(SB)) telescopes (
$\sim$ 12$\mu m$ $\Delta$E, 300$\mu m$ E) mounted in one arm of the  91.5
cm scattering chamber. Typical solid angle subtended by each detector was
$\sim$0.3  msr.  A monitor detector ($\sim$300$\mu$ Si(SB)) was placed in
the other arm of the scattering chamber for  normalisation  purpose.  The
telescopes  were calibrated using elastically scattered $^{16}$O ion from
Au target and $\alpha$-particle from (Th-$\alpha$) source. Typical energy
resolution obtained for the elastic $^{16}$O peak was $\sim $375 keV.

Inclusive  energy  distributions  for various fragments (3$\leq Z \leq$7)
were measured in the angular  range  15$^\circ$-115$^\circ$.  The  energy
spectra  of  the  emitted  fragments (3$\leq Z \leq$7) have been shown in
fig. 1 for $\theta_{lab} = 20 ^{\circ}$. The  systematic  errors  in  the
data,  arising from the uncertainties in the measurements of solid angle,
target thickness and the  calibration  of  current  digitizer  have  been
estimated to be $\approx$ 10\%.

\begin{figure}

\vspace{-3.cm}

\centering
{\epsfig{figure=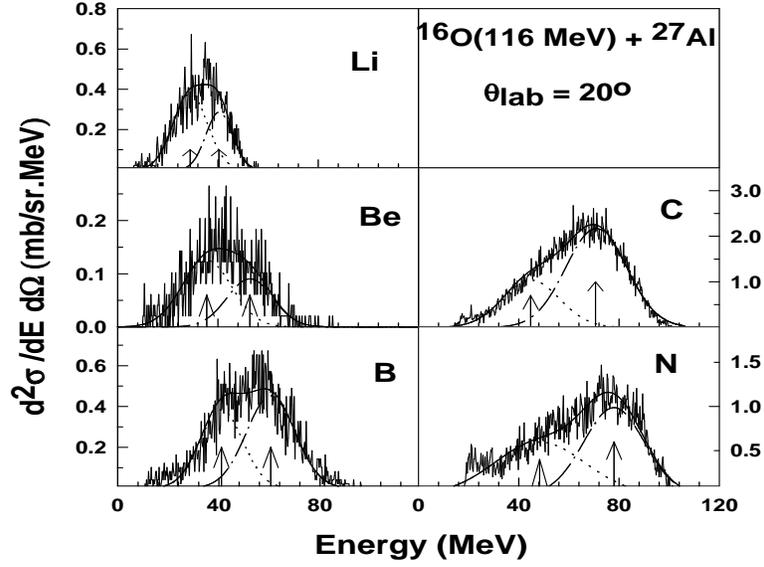,width=10.5cm,height=12.0cm}}

\vspace{-2.0cm}

\caption{  Energy  spectra  of different fragments obtained at 20$^\circ$
for the $^{16}$O+$^{27}$Al reaction (solid lines).  Dotted  and  dash-dot
lines  are  the  Gaussian fit to FF and DI components, respectively. Left
and  right  arrows  correspond  to   the   centroids   of   FF   and   DI
components,respectively.} \label{lico} \end{figure}

 It is evident from Fig.~\ref{lico} that the shapes of the energy spectra
of  heavier  fragments  (viz. C, N) are quite different from those of the
lighter fragments viz, Li and Be.  It  is  mainly  due  to  variation  of
relative contributions of DI and FF processes for different fragments. We
adopt  the following prescription for the estimation of FF, DI components
present in the spectra. The energy spectra of different fragments at each
angle  have been fitted with two Gaussian functions in the following way.
In the first step, the FF contributions have been obtained by fitting the
energy distributions with a Gaussian  having  centroid  at  the  energies
obtained  from  Viola  systematics  \cite{viola,beck96}  of total kinetic
energies(TKE) of mass-symmetric  fission  fragments  duly  corrected  for
asymmetric  factor\cite{beck98}  . The width of the Gaussian was obtained
by fitting the lower energy tail  of  the  spectra,  assuming  it  to  be
originating  from  purely  FF  process.  The  FF  component of the energy
spectrum thus obtained is then substracted from the full energy spectrum.
In the next step, the DI component is obtained by fitting the substracted
energy spectra with a second Gaussian. The above procedure is illustrated
in Fig.~\ref{lico} for the fragments ranging from Li to N at  20$^\circ$.
The dotted line in Fig.~\ref{lico} shows the contribution of FF component
and  the  dashed  dotted line shows the contribution of DI component. The
solid line shows the sum total contribution of both FF and DI components.
In each spectrum the arrow at lower energy corresponds to the centroid of
the Gaussian for the FF component obtained from Viola systematics and the
arrow at higher energy corresponds to the centroid of  the  Gaussian  for
the DI component.

\begin{figure}

\vspace{-1.cm}

\centering
{\epsfig{figure=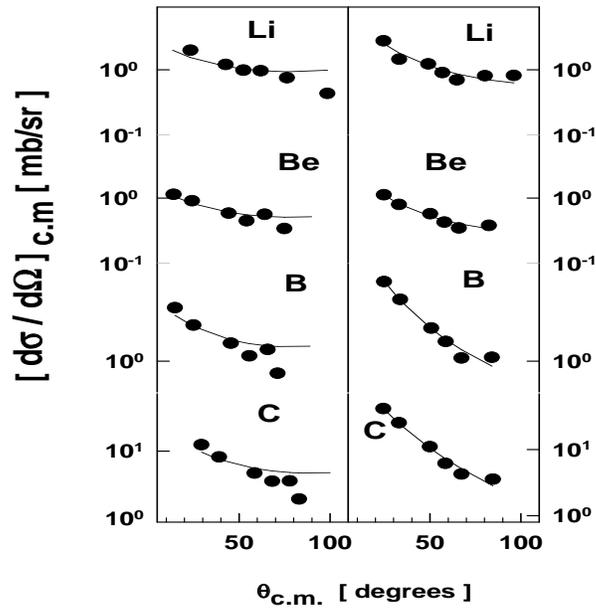,width=12.5cm,height=12.9cm}}

\vspace{-1.cm}
\caption{  Centre of mass angular distributions of different fragments FF
component ( left ) and DI component ( right ).}
\label{ffsig3}
\end{figure}

\begin{figure}

\vspace{-10.0cm}

\centering
{\epsfig{figure=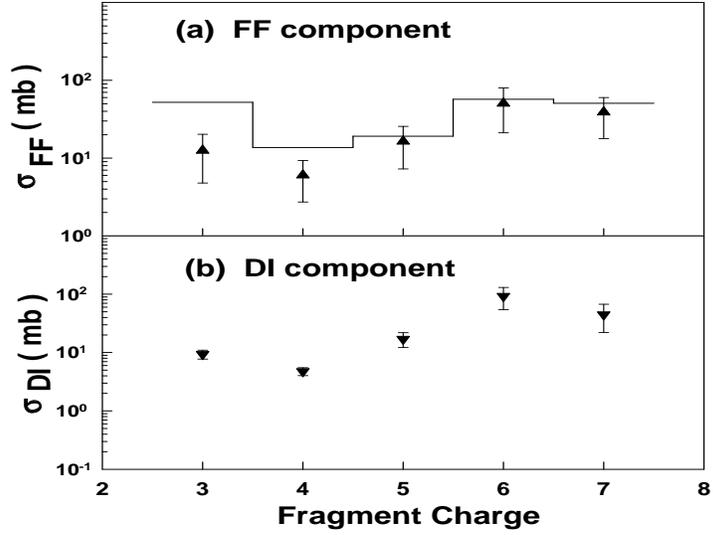,width=10.cm,height=11.0cm}}

\vspace{-1.cm}

\caption{   Fusion-  Fission  Fragment  emission  cross-sections.  Filled
circles and solid lines correspond to  the  experimental  and  calculated
results ( EHFM ), respectively.} \label{sigtot} \end{figure}

The  FF  and the DI components of the fragment angular distributions have
been obtained by integrating the respective energy distributions obtained
in the  manner  discussed  above.  The  centre  of  mass  (c.m.)  angular
distributions  of FF components of the fragments ( 3$\leq$Z$\leq$6 ) have
been  displayed  as  a  function  of  c.m.   angle   $\theta_{c.m.}$   in
Fig.~\ref{ffsig3} ( left ). The transformation from the laboratory system
to  the  c.m.  system  has  been  done  with the assumption of a two body
kinematics averaged over the whole range of c.m. angles. The solid  lines
correspond  to  1/sin$\theta_{c.m.}$  function  . It is clear that the FF
angular   distributions    for    different    fragments    follow    the
1/sin$\theta_{c.m.}$  type  of dependence, which is characteristic of the
decay of a fully  equilibrated  system  (  fusion-  fission  of  compound
nucleus  and/or  orbiting dinuclear system ). Total elemental yield of FF
component of the fragment emission cross-sections has been compared  with
the  theoretical  estimates  of  the  same  obtained  from  the  Extended
Hauser-Feshbach   Method   (EHFM)   \cite{Matsuse97,beck98}.   The   EHFM
calculations  have  been  performed  by using a critical angular momentum
value of $l_{crit} = 34\hbar$ and a neck parameter  consistent  with  the
systematics  given in ref.\cite{beck98}. The calculated fragment emission
cross-sections are shown in Fig.~\ref{sigtot}(a) as solid  histogram  and
compared with the experimental estimates of the same (filled circles). It
is  seen  from  the  figure  that the theoretical predictions are in fair
agreement with the experimental results. Therefore, it  may  be  inferred
that the extraction of the FF component of the fragment spectra following
the prescription described above, (using the Gaussian with centroid given
by the Viola systematics) is quite successful.

The  c.m.  angular  distributions  of  DI  components  of the fragments (
3$\leq$Z$\leq$6 ) have  been  displayed  as  a  function  of  c.m.  angle
$\theta_{c.m.}$  in  Fig.~\ref{ffsig3}  (  right)  .  A rapid fall of the
angular distribution than predicted by 1/sin$\theta_{c.m.}$  distribution
indicates a shorter life time of the composite system. Such lifetimes are
incompatible  with the formation of an equilibrated compound nucleus, but
may still reflect significant  energy  damping  within  a  deep-inelastic
mechanism.  From  the  measured forward peaked angular distribution it is
possible to estimate the life time of the intermediate di-nuclear complex
using a diffractive Regge-pole model \cite{Mikumo80,beck98}. The  angular
distributions are fitted with the following expression

\begin{equation}
                 d\sigma/d\Omega=                  (
C/sin\theta_{c.m.})(e^{-\theta_{c.m.}/\omega t})
\end{equation}

\noindent
and   the   fit   to  the  DI  component  of  the  spectra  is  shown  in
Fig.~\ref{ffsig3}(right).  This  expression  describes  the  decay  of  a
di-nucleus  rotating  with  angular velocity $\omega$=$\hbar l/\mu R^{2}$
where $\mu$ represents the reduced mass of the system , $l$  its  angular
momentum  (which  should  fall  somewhere  between  grazing ($l_{g}$) and
critical ($l_{cr}$) angular momentum), R represents the distance  between
the  two  centres  of  the di-nucleus and $t$ is the time interval during
which the two nuclei remain in  a  solid  contact  in  the  form  of  the
rotating  di-nucleus.  Small  values of the 'life angle' $\alpha(= \omega
t)$ lead to forward peaked angular distributions,  associated  with  fast
processes,  whereas  large  values  of  $\alpha$ , associated with longer
times    as    compared    to    the    di-nucleus    rotation     period
$t$(=2$\pi$/$\omega$),   are  consequently  associated  with  long  lived
configurations and lead to more isotropic angular distributions.  In  the
limiting  case  of  very  long-lived  configurations  , the distributions
approach  a  d$\sigma$/d$\Omega$  $  \propto$   (   1/sin$\theta_{c.m.})$
dependence.  The  time scales thus obtained are given in Table~\ref{tab1}
for different fragment charge Z. As found in a previous study  by  Mikumo
et.  al. \cite{Mikumo80} for the same reaction at 88 MeV, the time scales
decrease as the fragment charges increase. This is expected  because  the
heavier  fragments  (  nearer  to the projectile ) require less number of
nuleon transfer and therefore less time; on the other hand  the  emission
of  lighter  fragments  requires  exchange of more number of nucleons and
therefore longer times. Our quantitative analysis is  consistent  with  a
recent  qualitative  study of formation time in light heavy ion reactions
\cite{Szanto96}.

\begin{figure}

\vspace{-.8cm}

\centering
{\epsfig{figure=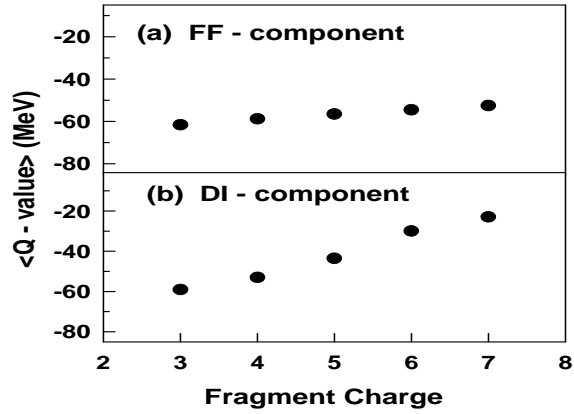,width=12.5cm,height=12.0cm}}

\vspace{-3cm}

\caption{  $<$Q$>$  value  for  FF  and  DI  components  of  different
fragments.}
\label{qval}
\end{figure}

In  Fig.~\ref{qval},  the optimum Q values ($<$Q$>$) generated for FF and
DI components of different fragments have been plotted as a  function  of
fragment  charge  Z  for  a typical angle $\theta_{lab}$=20$^0$. From the
figure, it is observed that $<$Q$>$ for FF  fragments  is  more  negative
than  for DI components and does not show much variation . This is due to
the fact that for FF process,  energy  relaxation  is  complete  and  the
system  is  fully  equilibrated. The small variation in $<$Q$>$ is due to
the variation  of  mass  asymmetry  of  the  fragments.  In  case  of  DI
component,  the large variation in $<$Q$>$ values is due to the different
extent of energy damping corresponding to variation in the degree of mass
transfer. Similar results have been observed at lower  incident  energies
for  the  same  reaction \cite{shapi2,Harris77,Mikumo80,Sato83}. However,
the ($<$Q$>$) for each  fragment  is  much  higher  than  those  ovserved
earlier  (\cite{Harris77,Mikumo80,Sato83}  at  lower projectile energies.
Such energy dependence of ($<$Q$>$) may be due to long life time  of  the
di-nuclear system.

The  total  fusion-fission ($\sigma_{FF}$) and the total deep-inelastic (
$\sigma_{DI}$) cross-sections for different fragments have been  obtained
by integrating the energy distribution of fusion-fission component and DI
component,  respectively  (  as discussed earlier) over the corresponding
energies and over the measured angles. The  cross-section  thus  obtained
for   different   fragments   have  been  displayed  in  fig  3a  and  3b
respectively, as a function of fragments Z. Total uncertainities  in  the
estimation of $\sigma_{FF}$ due to experimental threshold and the limited
angular  range  of  the  data  have  been  shown  by  the error - bars in
Fig.~\ref{sigtot}. It has been found that a large fraction  of  C  and  N
cross section is due to DI mechanism.

In   conclusion  we  have  measured  the  inclusive  double  differential
cross-sections  for  fragments  emitted  in  the  reactions  $^{16}$O   +
$^{27}$Al  at  $E_{lab}  =  116$  MeV . Total emission cross-sections for
various  fragments  have  been  deduced  from  the  double   differential
cross-section data. The shapes of the energy spectra of lighter fragments
e.g.  Li and Be, are quite different from those of the heavier fragments.
This may be due to additional contributions of QE and  DI  components  in
the  spectra  of  heavier  fragments.  It  is  observed  that the angular
distribution of the FF  component  for  different  fragments  follow  the
1/sin$\theta_{c.m.}$  type  of dependence, which is characteristic of the
fission like ( fusion-fission and/or orbiting) decay of  an  equilibrated
compound nucleus. Moreover, the predicted frgment emission cross-sections
using  the  Extended  Hauser-Feshbach  Method agree quite well with those
coming from the FF component. However, the c.m. angular distributions  of
DI components do not follow 1/sin$\theta_{c.m.}$ type of dependence . The
angular  distributions  of  the  DI components have been fitted using the
function (  C/sin$\theta_{c.m.})(e^{-\theta_{c.m.}/\omega  t})$  and  the
time  scale  for the emission of different fragments have been estimated.
The emission time is found to decrease as the fragment  charge  increases
which  is  expected  to  be  true intuitively. The total fusion-fission (
$\sigma_{FF}$)  and  deep-inelastic  ($\sigma_{DI}$)  cross-sections  for
different   fragments  have  been  obtained  by  integrating  the  energy
distribution of fusion-fission component and DI component ( as  discussed
in  the  text  )  over  the  corresponding energies and over the measured
angles. Although a large fraction of C and N cross section is due  to  DI
mechanism,  the  FF  process  is  found  to  be rather competitive in the
$^{16}$O + $^{27}$Al reaction, in agreement with the previous studies  of
the neighbouring $^{16}$O + $^{28}$Si system \cite{Szanto97}.

The  authors thank the accelerator operation staff of VECC for the smooth
running of the machine and staff of the target and detector  laboratories
for  providing the targets and the Si detectors. They are thankful to
C. Beck , for his constructive comments.  
One  of  the  authors (KM) acknowledges with thanks the
financial support received from C.S.I.R., India.

\vspace{7.0cm}
\centering
\begin{table}
\caption{Life  times  of  the  dinuclear  systems  for  different emitted
fragments.}    \begin{tabular}{cccccc}    &&&&&\\    &&&&&\\     Fragment
&\hspace{.5cm}Li    &   \hspace{.5cm}Be&\hspace{.5cm}B   &\hspace{.5cm}C&
\hspace{.5cm}N\\ &&&&&\\ &&&&&\\ \hline &&&&&\\ &&&&&\\ Time& &&&&\\
  (10$^{-22}$sec)&4.7&3.5&1.9&1.1&0.8\\
&&&&&\\
\end{tabular}
\label{tab1}
\end{table}

\end{document}